\documentclass[epj]{webofc}
\usepackage[varg]{txfonts}   
\usepackage{lineno}
\usepackage{lineno}
\usepackage{wrapfig}
\begin{document}
%
%
\title{Latest results on dark matter searches with H.E.S.S}
%
%

\author{Lucia Rinchiuso\inst{1}\fnsep\thanks{\email{lucia.rinchiuso@cea.fr}} 
        on behalf of the H.E.S.S. Collaboration}

\institute{IRFU, CEA, 
Universit{\'{e}} Paris-Saclay, F-91191 Gif-sur-Yvette, France}

\abstract{%
The nature of Dark Matter (DM) is one of the most debated questions of contemporary physics. Ground-based arrays of Cherenkov telescopes such as the High Energy Spectroscopic System (H.E.S.S.) search for DM signatures through the detection of Very-High-Energy (VHE, $E > 100$~GeV) gamma-rays. DM particles could self-annihilate in dense environments producing VHE $\gamma$-rays in the final states that could be eventually detected by H.E.S.S.. The H.E.S.S. observation strategy for DM search focuses towards the Galactic Centre (GC) region and nearby dwarf galaxy satellites of the Milky Way. The GC dataset provides the most stringent constraints on the DM annihilation cross section in the mass range 300 GeV - 70 TeV. Searches have been carried out towards classical and ultra-faint dwarf galaxies to test specific heavy DM models. 
The latest results towards the GC and dwarf galaxies are shown.
}
\maketitle
\section{The H.E.S.S. experiment}
\vspace{0.3cm}
The High Energy Spectroscopic System (H.E.S.S.) is an array of Imaging Atmospheric Cherenkov Telescopes (IACTs) located in Namibia. During the first phase (H.E.S.S. I) the experiment consisted of four telescopes of 12 m diameter. The phase 2 (H.E.S.S. II) has started with the addition of a 28 m diameter telescope in the middle of the array to lower the energy threshold down to a few tens GeV. H.E.S.S. can detect Very-High-Energy (VHE, $\gtrsim$ 100 GeV) $\gamma$-rays, that interact in the atmosphere creating electromagnetic showers. The showers then produce Cherenkov light that is detected by the telescopes situated in the light cone. The position and energy of the primary $\gamma$-ray can be reconstructed from the shower images recorded in the camera. The $\gamma$-rays are separated from the hadronic background based on image shape parameters of the shower. The H.E.S.S. array has a Field of View (FoV) of $5^\circ$, angular resolution per $\gamma$ better than $0.1^\circ$ and energy resolution $10\%E$. For observations at zenith towards a point-like source, it reaches a flux sensitivity 1\% of the Crab flux in 25 hours.
\section{Indirect dark matter search}
\subsection{Dark matter annihilation}
IACTs can probe Dark Matter (DM) indirectly, {\it i.e.} through the detection of VHE $\gamma$-rays produced in the final state of 
self-annihilation of DM. The expected $\gamma$-ray flux from DM annihilation writes:
$$\frac{d\phi_\gamma}{dE}=\frac{1}{4\pi}\frac{\langle\sigma v\rangle}{2m_{\rm DM}}\sum_{\rm i}Br_{\rm i}\frac{dN_{\rm i}}{dE}\times J(\Delta\Omega)\, ,$$
where $m_{\rm DM}$ is the DM mass, $\langle\sigma v\rangle$ the thermally-averaged velocity-weighted annihilation cross section, $Br_{\rm i}dN_{\rm i}/dE$ the spectrum of the annihilation channel $i$ weighted by its branching ratio $Br_i$. The J-factor $J$ describes the density distribution of DM in the target. We refer to the prompt photon annihilation channel as {\it $\gamma$-line}, and the annihilation channels with photons due to decay or hadronization of vector bosons, quarks and leptons in the final state as the {\it continuum}. 
The expected DM signal is large in environments with large $J$. The signal region, referred to as the {\it ON region}, is defined as a disk around the target, while the background is measured in signal-free or lower-signal regions in the same FoV ({\it OFF region}).
\subsection{Likelihood analysis technique}
The $\gamma$-ray events are measured in the {\it ON} and {\it OFF} regions. If no significant excess is observed between  the {\it ON} and {\it OFF} regions a Log-Likelihood Ratio Test Statistic (LLR TS) is performed to set 95\% confidence level (C.L.) limits on $\langle\sigma v\rangle$ for a fixed DM density profile and a specific annihilation channel. The likelihood function used in H.E.S.S. in the latest analyses is binned in two dimensions (2D): energy and space. This choice is made to take advantage of the peculiar spectral shape and spatial distribution of DM signal compared to the power-law-like and spatially-isotropic background. The likelihood function contains two Poisson terms for the {\it ON} and for the {\it OFF regions}, respectively:
$$\mathcal{L}_{\rm ij}=Poiss(N_{\rm ON,ij},N_{\rm S,ij}+N_{\rm B,ij})\times Poiss(N_{\rm OFF,ij},N_{\rm S,ij}'+\alpha_{\rm i}N_{\rm B,ij}).$$
$N_{\rm ON}$ and $N_{\rm OFF}$ are the measured events, $N_{\rm S}$ and $N_{\rm S}'$ the expected signal in the {\it ON} and {\it OFF regions}, respectively, $N_{\rm B}$ is the expected background in the {\it ON region} and $\alpha$ is the ratio between the size of the {\it OFF} and {\it ON regions}. 
\section{The Galactic Center region}
\begin{wrapfigure}[20]{l}{0.5\textwidth}
\vspace{-0.5 cm}
\centering
\includegraphics[width=0.56\textwidth]{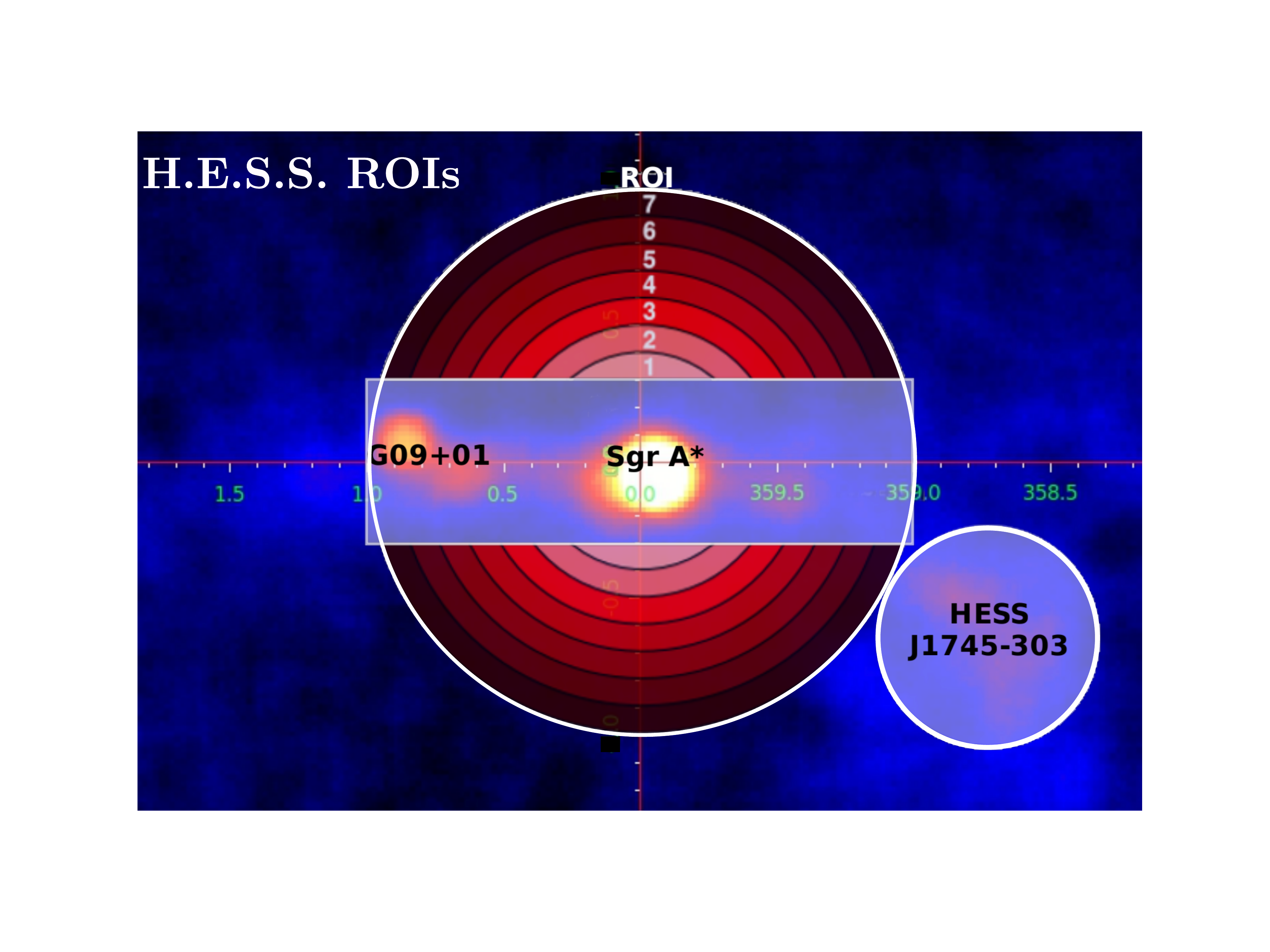}
\vspace{-1.3 cm}
\caption{Galactic Center sky map in Galactic longitude and latitude. The RoIs are shown in red and the excluded regions as shaded areas. The standard VHE $\gamma$-ray sources Sgr A*, G09+01 and HESS J1745-303 are drawn.}
\label{fig:GCmap}
\end{wrapfigure}
The region around the Galactic Center (GC) is expected to harbor a large amount of DM. For this reason and due to its proximity it is an optimal target for DM detection. 
In addition H.E.S.S. is in a preferential position to observe it being situated in the Southern hemisphere. H.E.S.S. I has collected about 250 hours of observations towards the GC in ten years. Due to the large dataset and the telescope pointing strategy an {\it ON} region up to one degree in radius can be defined around the target. It is then split into seven concentric subregions (RoI) with inner radii from $0.3^\circ$ to $0.9^\circ$, that make up the spatial bins for the likelihood function.\\ 
The {\it OFF region} is defined for each RoI and each observation 
as the region symmetric to the {\it ON} one with respect to the pointing position of the observation. Thus, the {\it ON} and {\it OFF regions} have the same solid angle size ($\alpha=1$) and are in the same FoV: they are observed under the same conditions, no acceptance correction is needed due to azimuthal symmetry. 
However, the GC region harbors several VHE $\gamma$-ray sources that must be rejected for the DM search analysis.
Fig.~\ref{fig:GCmap} shows a GC sky map with overlaid the {\it ON region} split in seven RoIs. The VHE $\gamma$-ray sources HESS J1745-290 coincident with the supermassive black hole Sagittarius A*~\cite{bib:1745-290}, G09+01~\cite{bib:G0901} and HESS J1745-303~\cite{bib:1745-303} are shown in the map.  
In the DM search analysis masks (shaded areas in Fig.~\ref{fig:GCmap}) are used over these sources and the Galactic plane diffuse emission to avoid contamination from $\gamma$-rays produced in standard astrophysical processes. These areas are excluded both in the {\it ON} and {\it OFF regions} in order to maintain the same solid angle size. \\
No significant excess is observed in the {\it ON region} versus the OFF region. So the 2D-likelihood analysis is performed in the {\it continuum}~\cite{bib:continuum} and {\it $\gamma$ line}~\cite{bib:line} channels, assuming an Einasto DM density profile. The LLR TS is applied to test the hypothesis of presence of DM vs the no-DM hypothesis and to obtain the 95\% C.L. upper limits on the annihilation cross section. The results for the {\it $\gamma$-lines} search are shown in Fig.~\ref{fig:limits} left (red dots) together with the expected limits (black solid line), their containment bands at $1\sigma$ and $2\sigma$ (green and yellow boxes, respectively) and the previous H.E.S.S. limits (blue squares)~\cite{bib:GC13}. Thanks to the novel 2D-likelihood technique, the more performing raw data analysis~\cite{bib:MdN} and the ten-year data set, the constraints are improved of a factor about 6 at 1 TeV, reaching $\langle\sigma v\rangle \simeq 4\times10^{-28}$~cm$^3$~s$^{-1}$. The H.E.S.S. limits become competitive with Fermi-LAT~\cite{bib:Fermi} at a few hundred GeV. The two experiments together put the strongest constraints so far on $\langle\sigma v\rangle$ in the GeV-TeV DM mass range.
\section{The dwarf galaxies satellites of the Milky Way}
\begin{figure}[htb]
\begin{center}
\vspace{-0.3 cm}
\includegraphics[width=0.45\textwidth]{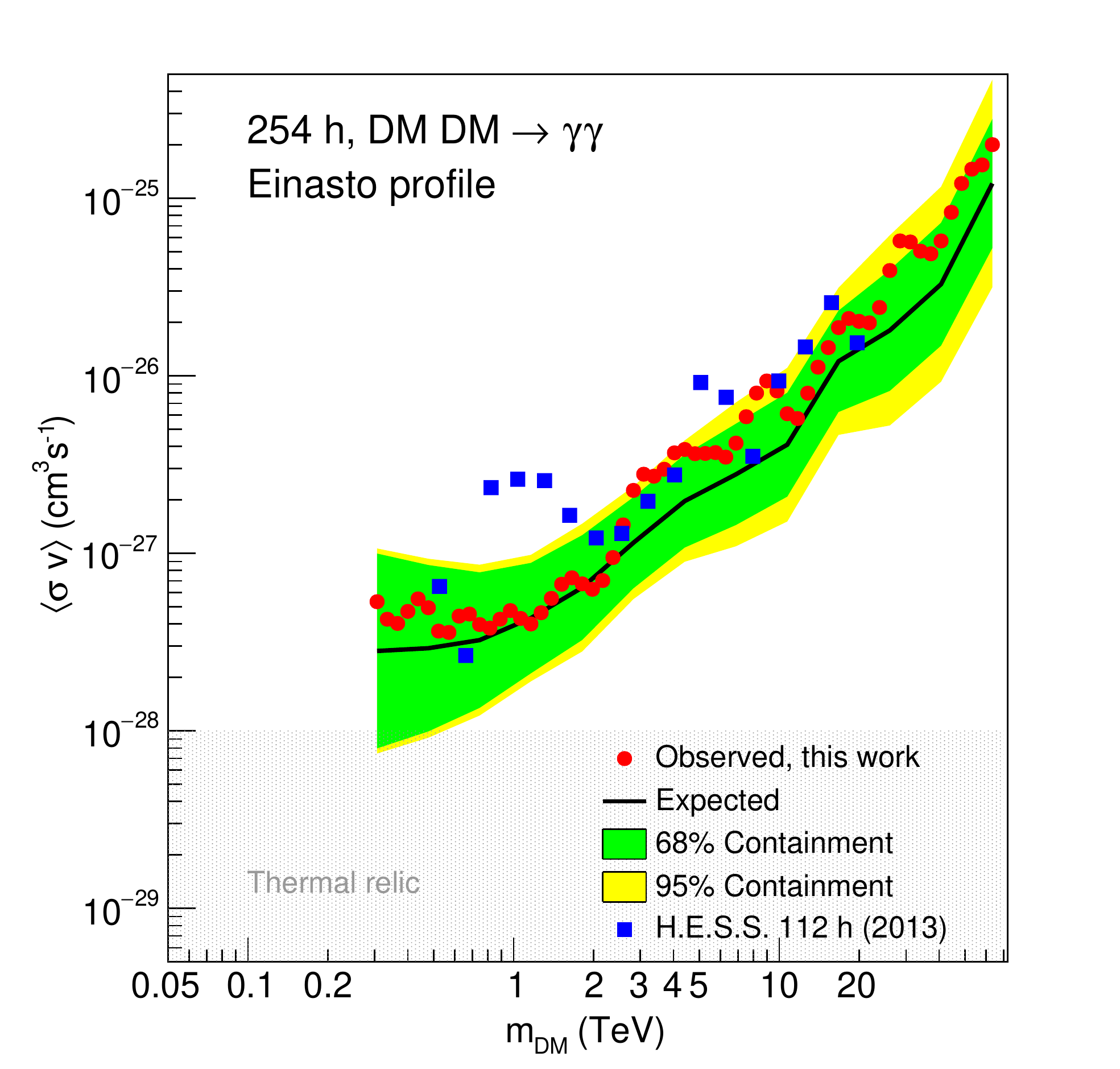}
\includegraphics[width=0.45\textwidth]{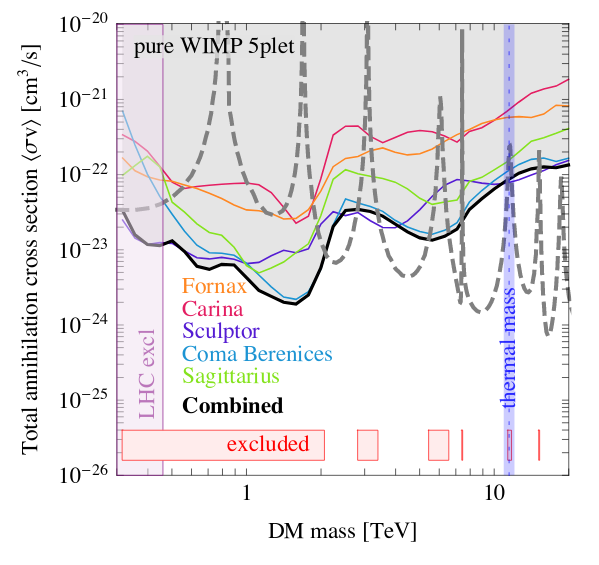}
\caption{95\% C.~L. upper limits on $\langle \sigma v \rangle$ as a function of $m_{\rm DM}$ . {\it Left:} Limits  for the {\it $gamma$-line} derived from H.E.S.S. observations taken over ten years (254~h live time) of the inner 300 pc of the GC region. Observed limits (red dots) and mean expected limit (black solid line) are shown together with the $1\sigma$ (green band) and $2\sigma$ (yellow band) containment bands. {\it Right:} Limits for the 5plet towards dwarf galaxies are shown for single galaxy observation and for their combination (black solid line). The predicted cross section (gray dashed line), thermal mass (blue band) and excluded masses (red boxes) are represented.}
\label{fig:limits}
\end{center}
\end{figure}

Satellite dwarf Spheroidal galaxies (dSph) of the Milky Way are among the most DM dominated objects in the Universe. 
In addition, they do not show VHE $\gamma$-ray emission from standard sources, e.g. there is no hints for recent star formation. Thus, they are expected to have a large signal-to-noise ratio and are good targets that could provide an unambiguous detection of a DM signal. H.E.S.S. I has accumulated a total of about 130 hours of observations towards Carina (23 h), Coma Berenices (11 h), Fornax (6 h), Sculptor (12 h) and Sagittarius (85 h). On these targets the {\it ON region} measures $0.2^\circ$ ($0.3^\circ$ for Coma Berenice) accordingly to the DM density profile and is split in RoIs of width $0.1^\circ$. The {\it OFF regions} are defined for each RoI and each observation with the {\it multiple OFF} technique ($\alpha>1$).\\
No significant excess is observed in the {\it ON region} with respect to the {\it OFF}, in any of the dSphs. The 2D 
LLR TS is applied to compute  95\% C.L. upper limits on the different targets  for the {\it $gamma$-line} signal. The best combined constraint of $5\times10^{-25}$~cm$^3$~s$^{-1}$ is obtained at 1 TeV. A further analysis has been conducted on specific pure WIMP (Weakly Interacting Massive Particle) models. Limits have been computed for the Minimal DM fermionic triplet (3plet) and quintuplet (5plet)~\cite{bib:SUSY} using recent calculations~\cite{bib:dSph}. The full WIMP spectrum is used in the analysis, including $W^+W^-$ tree-level and $ZZ$, $Z\gamma$ and $\gamma\gamma$ one-loop contributions. 
Fig.~\ref{fig:limits} right~\cite{bib:dSph} shows the observed limits on the 5plet for single dSph observation and the combination of the five dSph observation (black solid line). The predicted cross section (gray dashed line) and thermal $m_{\rm DM}$ (blue band) are shown. A significant $m_{\rm DM}$ range is excluded for the 5plet (red boxes).
%
\section{Prospects}
H.E.S.S. is pursuing a DM program 
focused on the search for DM annihilating 
towards the GC and dSph. 
A new strategy has been developed in the GC region for further DM studies. A survey at higher latitudes from the GC is ongoing with the full telescopes array. These new observations are expected to provide more than doubled statistics and give the possibility to increase the size of the {\it ON region}, pushing the limits to smaller $\langle\sigma v\rangle$ values. Two projects have also been started for the dSph. An observation campaign of some of the recently-discovered DES dSph~\cite{bib:DES} is ongoing. A project has begun among the H.E.S.S., MAGIC and VERITAS collaborations for a common effort to jointly search for DM towards dSph exploiting the complementarity of the different instruments and available datasets for more powerful analyses.
%
%

\end{document}